# Modeling based screening for optimal carrier selective material for Si based solar cells

Nithin Chatterji[1,**], Aldrin Antony[2], and Pradeep R. Nair[1,*]
[1]Dept. of Electrical Engineering,
[2]Dept. of Energy Science and Engineering,
Indian Institute of Technology Bombay, Mumbai, India
email id- *prnair@ee.iitb.ac.in ,**nithin@ee.iitb.ac.in

*Abstract* – Carrier selective (CS) silicon solar cells are increasingly explored using a variety of different materials. However, the optimum properties of such CS materials are not well understood. In this context, through detailed analytical and numerical modeling, here we provide several interesting insights on the efficiency tradeoff with CS material properties. First, we show that perfect band alignment is a desirable feature only if the interface is devoid of any trap states. Otherwise, a band offset of around 0.2eV-0.4eV provides sufficient band bending to reduce the effect of interface recombination, thus improving the performance. Surprisingly, the interface passivation quality for the minority carrier extraction layer is found to be far less demanding than that for the majority carrier extraction layer. Additionally, doping density and dielectric constant of CS layers have a similar effect as band offset on solar cell performance. Our results have obvious implications toward the selection of appropriate materials as carrier selective layers and hence are of broad interest to the community.

## I. INTRODUCTION

Silicon based heterojunction devices with carrier selective contacts are increasingly explored as a cost effective alternative for the conventional diffused PN junction based c-Silicon solar cells[1]–[10]. Indeed, various techniques like atomic layer deposition[3], chemical vapor deposition[1], solution processing[6], etc., are explored for low temperature fabrication of carrier selective contact layers. Apart from the obvious advantages related to cost effectiveness, large band gap carrier selective layers can also reduce the parasitic absorption at the front end (as compared to the parasitic absorption loss in highly doped emitter of c-Si solar cells)[11]–[14]. Further, good conductivity can be achieved in these materials without intentional doping – which is usually a high temperature process which adds to the cost. As a result, different materials such as $TiO_2$[1]–[3], $LIF_x$[4], $KF_x$[5], PEDOT:PSS[6], $MoO_x$[7]–[10], $V_2O_5$[10], and $WO_3$[10] have been extensively studied recently. We note that there have been several modeling[15], [16] efforts as well to understand the device performance of various materials as carrier selective contacts with silicon.

In spite of the above exciting research, however, several critical aspects related to Si based carrier selective solar cells still remain unexplored or not well understood. For example - (a) how crucial is the band level alignment of the CS layer with c-Si, or rather is it essential to have the CS layer bands align with the respective bands of Silicon? (b) Given a band offset, what is the effect of CS/Si interface traps on the efficiency, and (c) with the above information, which pair of materials might be best suited for Si based solar cells. In this manuscript, we address the above mentioned topics through detailed modeling. For this, we first develop an analytical model to predict the functional dependence of device performance on CS material parameters (Section II). These predictions are then further refined through detailed numerical simulations (Section III). Curiously, our results indicate that the performance of the solar cell is the worst for the ideal case scenario in which there is no band discontinuity between the CS layer and Si (respective bands). Further, we provide a detailed map of efficiency vs. CS material parameters which could be of immense interest to the community to a-priori evaluate the performance of any pair of materials as carrier selective layers.

## II. ANALYTICAL MODEL

Figure 1 shows a schematic of the silicon heterojunction solar cell with carrier selective contacts along with the band level alignments. Here ESL and HSL denote the electron and hole selective layers, respectively. At the ESL/Si interface, hole transport from Si to ESL is blocked due to the large barrier between the valence bands of both materials. However, the transport of electrons at the same interface is more readily facilitated, even though there could be a band offset between the conduction bands. Note that a positive value for band offset indicates that the photo-generated carriers need to overcome a barrier to reach the corresponding transport layer, while a negative value for the offset indicates that the carrier injection from transport layer to silicon is limited by a potential barrier. Further there could be interface traps as well at the ESL/Si interface. Similarly, we assume perfect electron blocking characteristics and a valence band offset at the Si/HSL interface. Here we explicitly consider the performance trade-off of such solar cells as a function of transport barrier and interface traps. As such, many other factors could also influence the solar cell performance which includes the effective doping and thickness of carrier selective layers, nature of metal or TCO contact with the carrier selective layers, etc. With the aim of developing a coherent description of the various effects, here we make a few simplifying assumptions – (a) the contact layers are considered to be doped, (b) the metal or TCO contact with selective layers are assumed to be ohmic in nature, (c) over the barrier transport is assumed as the dominant transport mechanism at Si/CS layer interface, and (d) uniform density of traps at Si/CS layer interface. With these assumptions we first develop an analytical model to predict the device performance. Later, detailed numerical simulations (self-consistent solution of Poisson and carrier continuity equations) have been performed to further refine analytical predictions. The parameters used in this study are provided in the appendix A.

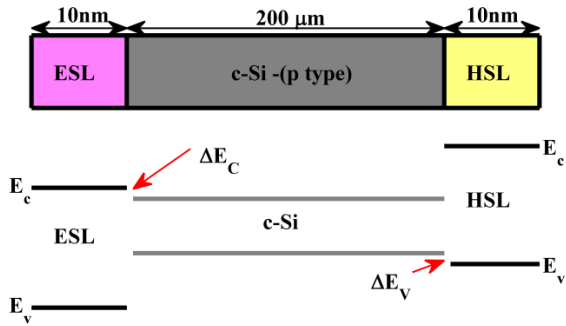

**Fig. 1**: Schematic (top) and energy level alignments (bottom) of a Si based carrier selective solar cell used to study the effect of band alignment on solar cell performance. In the numerical simulations, $\Delta E_c$ at the ESL/c-Si interface is varied from $-0.6eV$ to $+0.6eV$, while the barrier for holes is kept fixed ($2.28eV$). Similarly, the $\Delta E_v$ at the c-Si/HSL interface is varied from $-0.2eV$ to $+0.5eV$, while the barrier for electrons is kept fixed ($2.28eV$). Refer Table 1 for simulation parameters.

The material properties of CS layer that affect the electrostatics of the solar cell are (a) the band alignment with c-Si, (b) traps at the interface between CS layer and Si, (c) doping density and (d) the dielectric constant of CS layer. Note that these parameters are both material and process dependent. The parameters that dictate the efficiency of a solar cell are the open circuit potential ($V_{OC}$), short circuit current ($J_{SC}$), and the fill factor (FF)[17]. Of the above, the maximum achievable $V_{oc}$ is dictated by the detailed balance of carrier generation with various recombination mechanisms. Similarly, the $J_{sc}$ is a measure of the photo-generated carrier collection efficiency at short circuit conditions, while $FF$ is more influenced by the collection efficiency at maximum power point condition. Below, we develop an analytical model to predict the variation of the above three performance metrics as a function of CS material parameters. We first start with the effect at the ETL/Si interface with ideal conditions assumed for Si/HTL interface (i.e., zero band offset for holes, perfect electron blocking, and lack of any interface trap states).

*a) $V_{oc}$ estimation*: For p-type substrate, the maximum achievable $V_{oc}$ is given as.

$$V_{oc} = \frac{kT}{q}\ln\frac{G\tau_{eff}}{n_i} + \frac{kT}{q}\ln\frac{N_A}{n_i}, \quad (1)$$

where $G = 1.35 \times 10^{19} cm^{-3}s^{-1}$ is the uniform generation rate in c-Si corresponding to AM1.5 spectrum, and $N_A$ is the bulk doping density in c-Si. The effective minority carrier lifetime, $\tau_{eff}$[18], is given by

$$\frac{1}{\tau_{eff}} = \frac{1}{\tau_{bulk}} + \frac{s}{w_{Si}}, \quad (2)$$

where $\tau_{bulk}$ is the bulk lifetime which includes the SRH, radiative recombination, and Auger recombination mechanisms. Accordingly, for low levels of illumination we have $\tau_{bulk} = \frac{1}{\frac{1}{\tau_{SRH}}+BN_A+CN_A^2}$, where $B$ is the radiative recombination coefficient and $C$ is the Auger recombination coefficient. The parameter $s$ in eq. (2) denotes the surface recombination velocity at ESL/c-Si interface and $w_{Si}$ is the thickness of c-Si substrate. The SRH model[19] indicates that

$$s = \frac{R_s}{\Delta n_s} = \frac{\int_{E_v}^{E_c} \frac{n_s p_s - n_i^2}{\frac{n_s + n_{1s}}{c_{ps}} + \frac{p_s + p_{1s}}{c_{ns}}} D_{it} \, dE}{\Delta n_s}, \quad (3)$$

where $n_s$ and $p_s$ are the electron and the hole densities, respectively, at the ESL/c-Si interface under illumination and $\Delta n_s$ is the excess electron density at the ESL/c-Si interface in the presence of light. Under open circuit conditions, we have

$$n_s = n_{bulk} e^{\frac{\Delta \Psi}{kT/q}}, \quad (4)$$

$$p_s = p_{bulk} e^{-\frac{\Delta \Psi}{kT/q}}, \quad (5)$$

$$n_{bulk} = G \tau_{eff}, \quad (6)$$

where $n_{bulk}$, $p_{bulk}$ are the bulk electron and hole concentrations, respectively, in c-Si. $\Delta \Psi$ is the band bending in c-Si and is given as,

$$\Delta \Psi = \frac{N_D \varepsilon_{esl}(V_{bi} - V_{oc})}{N_A \varepsilon_{Si} + N_D \varepsilon_{esl}}. \quad (7)$$

$$V_{bi} = abs(\varphi_{esl} - \varphi_{Si}), \quad (8)$$

where $N_D$ is the doping density in ESL, $V_{bi}$ is the built in potential between ESL and c-Si, $\varepsilon_{ESL}, \varepsilon_{Si}$ and $\varphi_{esl}, \varphi_{Si}$ are the dielectric constants and work functions of ESL, c-Si respectively.

Equations (1) – (8) describe the electrostatics of the device and a self-consistent solution of the same predicts the variation of open circuit voltage. It can be used to estimate the effect of band offset (which affects the $\varphi_{esl}$, see eq. 8), effect of interface traps, doping density, and dielectric constant of CS layers on the device performance. The values of various parameters used in this study are provided in Table 1. As our aim is to study the effect of the above mentioned parameters in the extraction of photo-generated carriers, the barrier for blocking the carriers at the respective selective layers is kept large and fixed. For example, at the c-Si/ESL interface the $\Delta E_c$ is varied while the barrier for hole injection from c-Si to ESL is kept fixed (see Fig. 1). Since the number of critical parameters that affect the performance is large, we first focus on the effect of band offset and interface traps (see Fig. 2) and identify the physical mechanism that control the performance. Once this is achieved, the insights can be readily extended to other parameters like doping density and dielectric constant as well.

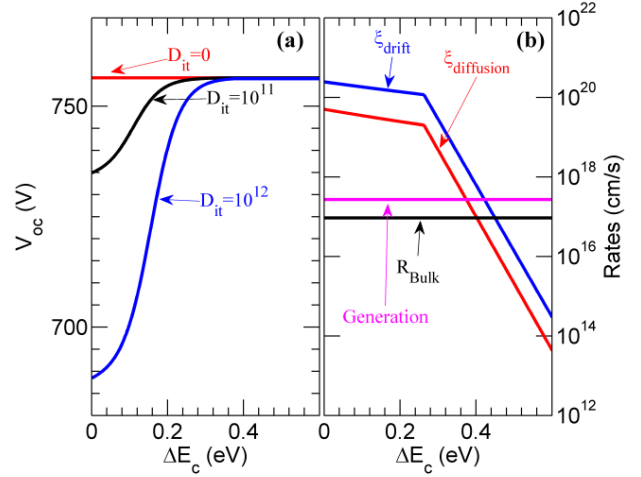

**Fig. 2**: The effect of the $\Delta E_c$ on Solar cell performance. (a) Variation of $V_{oc}$ with $\Delta E_c$. (b) Comparison of the photo-generated carrier extraction rates at ESL/Si interface (asymptotic analysis at short circuit conditions) with the bulk recombination rates.

Figure 2 (a) shows the variation of $V_{oc}$ with $\Delta E_c$ in the presence and absence of interface traps at ESL/c-Si interface. Here, we assume $\tau_{SRH} = 1ms$, $B_{RR} = 1.1 \times 10^{-14} \, cm^3 s^{-1}$, $C_{Auger(n,p)} \approx 10^{-31} cm^6 s^{-1}$ - these parameters are comparable to that of the best efficiency devices reported[20]. $V_{oc}$ is estimated by self consistent solution of equations (1)-(8) with the assumption that the band bending in Si is not larger than $2 \times (E_{Fp} - E_v) + \frac{kT}{q}$, which corresponds to strong inversion at ETL/c-Si interface.

There are several interesting insights in Fig. 2 (a). The $V_{oc}$ in the absence of interface traps depends only on the bulk lifetime and hence does not vary with $\Delta E_c$. However, in the presence of interface traps, $V_{oc}$ is minimum near $\Delta E_c = 0eV$ and it reaches a maximum for a particular conduction band offset $\Delta E_c$. This is a rather surprising result as $\Delta E_c = 0eV$ is supposed to give the best performance. This interesting result is further explored through detailed numerical simulations.

*b) Effect of $\Delta E_c$ on $J_{sc}$ and $FF$:* The analytical results in previous section indicate that the achievable $V_{oc}$ is expected to be the least for perfect band alignment due to increased interface recombination. The results also indicate that there should be a minimum band offset of around 0.2-0.4eV for optimal $V_{oc}$. However, this analysis still does not predict the trends for efficiency unless accounted for $J_{sc}$ and $FF$ – both depend on the collection efficiency of carriers at respective electrodes. The bias dependent collection efficiency can be estimated using the steady state continuity equation given by

$$\frac{J}{q} = Gw_{si} - R, \qquad (9)$$

where, $J$ is the current flowing out through the ESL, G is the carrier generation rate, $w_{si}$ is the substrate thickness, and $R$ is the net recombination in the device. Note that $\frac{J}{q}$ denotes the escape rate of electrons through the ESL. The diffusion component for escape rate is given by $\xi_{diffusion} = \frac{D}{w_{ESL}} n_s e^{-\frac{\Delta E_c}{kT}}$; where $D$ is the diffusion constant, $w_{ESL}$ is the thickness of ESL, and $n_s$ is the electron density at the ESL/c-Si interface under illumination. The drift term for escape rate is given by $\xi_{drift} = \mu \epsilon n_s e^{-\frac{\Delta E_c}{kT}}$, where $\mu$ is the mobility of electron in ESL, $\epsilon$ is the electric field in the ESL.

The carrier collection efficiency could be estimated by comparing the escape rates with the net generation rate. For this, accurate estimate is required for the interface carrier density $n_s$. Note that the $n_s$ predicted by eq. (4) is valid only in open circuit conditions and hence is not appropriate for short circuit conditions. Further, accurate estimates are required for the electric field in the ESL as well. As an asymptotic analysis, here we assume that $\epsilon = \frac{(V_{bi} - V_{app})_{ESL}}{w_{ESL}}$, which is the maximum possible electric field in ESL and hence the best chance for escape. Details on the estimation of bias dependent $n_s$ are provided in the appendix B.

The variation of asymptotic $\xi_{diffusion}$ and $\xi_{drift}$ at zero bias as a function of $\Delta E_c$ is given in Fig. 2 (b). The interface recombination term is absent in this case as $n_s p_s = n_i^2$ at $V_{app} = 0$. The figure predicts some interesting trends – (a) $J_{sc}$ is expected to be not affected for small $\Delta E_c$ as the collection probabilities (i.e., the drift and the diffusion terms) are much larger than the bulk recombination probability, and (b) the collection probabilities become comparable to that of recombination for larger $\Delta E_c$ (i.e., around 0.3-0.4eV), and hence $J_{sc}$ is expected to decrease as $\Delta E_c$ increases.

The variation of $FF$ with $\Delta E_c$ is rather difficult to anticipate. For small $\Delta E_c$, we expect that the $FF$ might follow the trends of $V_{oc}$ as predicted by the analytical relationship between $FF$ and $V_{oc}$[21]. For large $\Delta E_c$, the $FF$ decreases due to the reduction in collection efficiency of photo-generated carriers (trends similar to the short circuit conditions, see Fig. 2b).

The analysis in this section predicts that the performance of a carrier selective solar cell is not at its optimal value for perfect band alignment – indeed, the best device performance could be at an optimal $\Delta E_c$, which in turn could depend on the interface recombination as well. Indeed, the same model predicts that the performance improves with increase in the doping density and dielectric constant of the CS material (results not shown but these trends can be anticipated from eq. 7).

### III. NUMERICAL SIMULATIONS

To further explore the predictions from the analytical model, we performed detailed numerical simulations (self-consistent solution of Poisson and drift diffusion equations). Table 1 provides the list of parameters used in simulations. The effect of band discontinuity between c-Si and ESL in the presence and absence of traps is explored using numerical simulations. For this we first study the effect of conduction band offset $\Delta E_c$ at the ESL/Si interface while keeping $\Delta E_v = 0eV$ for HSL. Uniform density of interface traps was assumed at the ESL/c-Si interface. Later the effect of band offsets at Si/HSL interface is also discussed.

Figure 3a and 3b shows the energy band diagram near the ESL at short circuit conditions for $\Delta E_c = 0eV$ and $\Delta E_c = 0.4eV$, respectively. It is evident that the band bending in c-Si is more in the case for $\Delta E_c = 0.4eV$. This large band bending causes an inversion region near the c-Si/ESL interface. Fig. 3c and 3d show the variation in the carrier densities with bias at the ESL/c-Si interface in the c-Si edge for $\Delta E_c = 0eV$ and $\Delta E_c = 0.4eV$, respectively. Due to the increase in inversion charge for $\Delta E_c = 0.4eV$, the value of $n_s$ is more and the value of $p_s$ is less compared to the corresponding values for $\Delta E_c = 0eV$. These results indicate that the band bending at c-Si interface can significantly influence the interface recombination. For example, the minority carrier density, which dictates the rate of trap assisted recombination, is significantly lower for the case with large $\Delta E_c$. As a result, the $V_{oc}$ is expected to increase with $\Delta E_c$. Indeed, the effect of a band discontinuity between ESL/c-Si is very similar to that of field effect passivation[22] as one type of carriers is prevented from reaching the interface thus reducing the recombination. Accordingly, the interface recombination term, $\frac{n_s p_s c_{ps}}{(n_s + n_1)}$ is maximum and $V_{oc}$ is least for $\Delta E_c = 0 eV$. The effect of negative values of $\Delta E_c$ on the device electrostatics is provided in the appendix C.

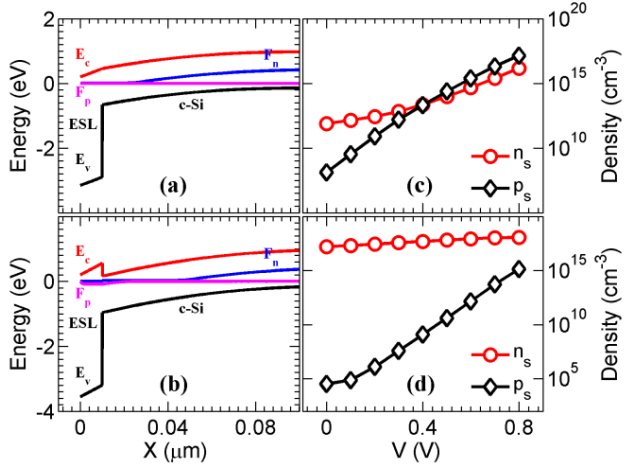

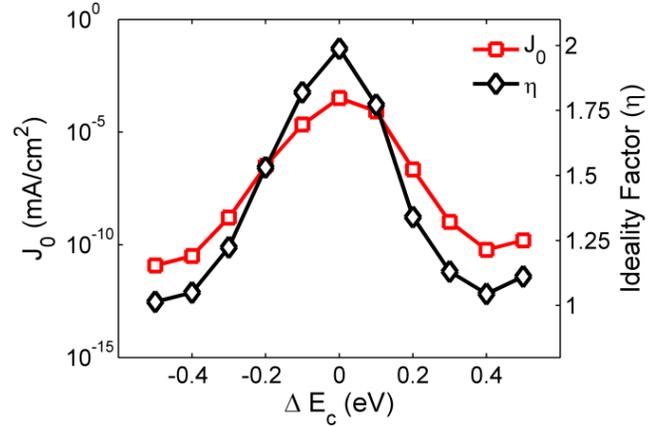

**Fig. 4**: The effect of $\Delta E_c$ on $J_0$ and ideality factor in the dark IV with interface traps ($10^{12} cm^{-2} eV^{-1}$). It indicates that $J_0$ and ideality factor improves as the band offset varies from $\Delta E_c = 0 eV$ conditions due to the reduction in interface recombination.

**Fig. 3**: The effect of $\Delta E_c$ on band bending and carrier densities at c-Si/ESL interface (numerical simulations, under illumination). Parts (a) and (b) show the energy band diagram for $\Delta E_c = 0 eV$ and $\Delta E_c = 0.4 eV$, respectively, at short circuit conditions. Parts (c, d) show the variation in interface carrier densities $n_s$ and $p_s$ with bias for $\Delta E_c = 0 eV$ and $\Delta E_c = 0.4 eV$, respectively.

The above insights are well supported by the trends related to dark IV characteristics as well. For example, the effect of $\Delta E_c$ on the dark IV characteristics in the presence of interface trap density of $10^{12} cm^{-2} eV^{-1}$ is shown in Fig. 4. Ideality factor is close to 2 at $\Delta E_c = 0 eV$ which correspond to significant recombination due to the interface traps. Ideality factor decreases with $\Delta E_c$ and reaches a value close to 1 near $\Delta E_c = \pm 0.4 eV$. This indicates a reduction in the detrimental effect of interface traps at larger values of $\Delta E_c$ as discussed before. Note that $J_0$ follows the trend of the ideality factor. $J_0$ has the maximum value at $\Delta E_c = 0 eV$ as a result of significant interface recombination - which is also confirmed by carrier densities, $n_s$ and $p_s$ in Fig. 3. We also notice that the $V_{oc}$ variation with $\Delta E_c$ can be accurately predicted with this ideality factor and $J_0$ and the details are provided in the appendix D.

Fig. 5 shows the effect of discontinuity in the conduction band between ESL and c-Si on the solar cell performance. To explore the details of the effect of interface traps, here we consider three cases: (a) a device with no interface traps, (b) a device with interface trap density $D_{it} = 10^{11} cm^{-2} eV^{-1}$ and (c) a device with interface trap density $D_{it} = 10^{12} cm^{-2} eV^{-1}$. Fig. 5a shows the variation in $V_{oc}$ with $\Delta E_c$. Here the effect of negative $\Delta E_c$ is also taken in to account. The results are similar to the predictions from the analytical model, however the absolute values of $V_{oc}$ are different in the two schemes. This discrepancy is due to the inaccuracy in analytical model related to the electrostatics of the device. In the absence of any interface traps, the $V_{oc}$ is independent of any band offset as $V_{oc}$ depends only on bulk carrier lifetime. As the interface trap density is increased, $V_{oc}$ decreases as the effective lifetime decreases. In the presence of traps the variation in $V_{oc}$ is almost symmetric with $\Delta E_c$. Minimum value of $V_{oc}$ is at $\Delta E_c = 0 eV$ and it increases in both the directions.

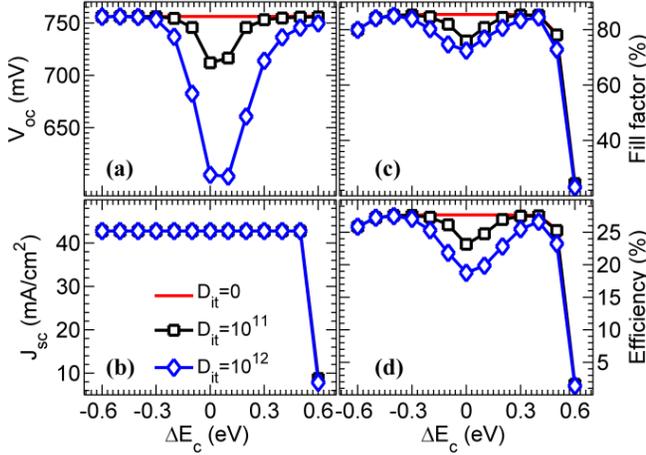

**Fig. 5**: The effect of the $\Delta E_c$ on performance parameters - (a) $V_{oc}$, (b) $J_{sc}$, (c) $FF$, and (d) efficiency of CS Si solar cells for different $D_{it}$ ($cm^{-2}eV^{-1}$). Note that the trends are broadly consistent with the analytical model and indicates that, surprisingly, the solar cell performance is ideal for a non-zero band offset

Figure 5a indicates that $V_{oc}$ varies almost linearly with $\Delta E_c$ before it saturates. This can be understood in simple terms as follows: As shown in Fig. 3, the effect of $\Delta E_c$ is to increase the electron concentration and decrease the hole concentration at the interface as compared to the bulk concentrations. Accordingly, the electron density in c-Si at the ESL/c-Si interface is significantly greater than the hole density at the interface and that the excess electron density at the interface is significantly greater than the dark electron density at the interface, i.e., $n_s \gg p_s, n_1$ and $n_s \sim \Delta n_s$. Assuming uniform distribution of traps, $s$ in eq.(3) can be approximated using eq.(4) as

$$s = \frac{D_{it} p_s c_{ps} \int_{E_v}^{E_c} dE}{n_s} = \frac{D_{it} p_{bulk} e^{-\frac{2\Delta E_c}{kT}} c_{ps} \int_{EV}^{Ec} dE}{n_{bulk}}. \quad (10)$$

If $\tau_{eff} \approx \frac{w}{s}$, then eq. (1) with eq. (10) predict a linear variation of $V_{oc}$ with $\Delta E_c$, as observed in Fig. 5a. Note that similar analysis is valid for negative values of $\Delta E_c$ as well.

Variation of $J_{sc}$ with $\Delta E_c$ is shown in Fig. 5b. As seen in the analytical model $J_{sc}$ is not affected till a particular value of $\Delta E_c$ is reached. After that the over the barrier transport of carriers to ESL decreases with increase in $\Delta E_c$. The collection of electrons is not typically affected with negative $\Delta E_c$, as the band bending at short circuit conditions is large enough (Fig. 4a, 4c), and hence there is no effect of interface traps on $J_{sc}$ for negative $\Delta E_c$.

Fig. 5c shows the variation in $FF$ with $\Delta E_c$. Without any interface traps, $FF$ is not affected till $\Delta E_c = 0.4eV$. Beyond $\Delta E_c = 0.4eV$, over the barrier transport of electrons from c-Si to ESL is affected at maximum power point conditions and hence gets reflected in $FF$. The results show that the presence of interface traps significantly affects the $FF$ for lower values of $\Delta E_c$. Specifically, under such conditions, the $FF$ follows the $V_{oc}$ trends for both the negative and positive values of $\Delta E_c$, as predicted by the empirical relationship connecting $FF$ with $V_{oc}$[21].

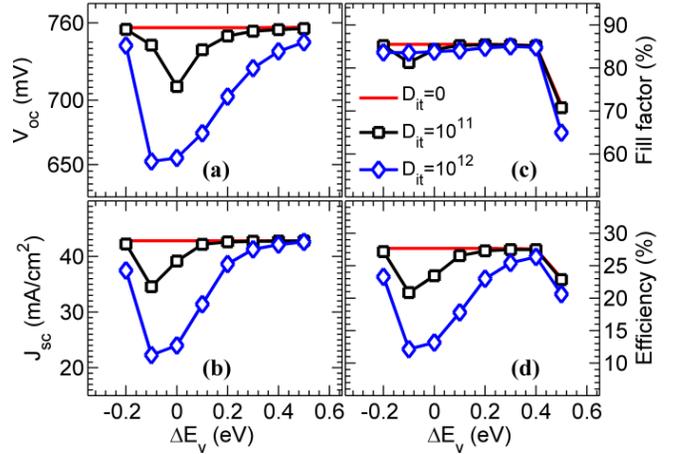

**Fig. 6:** The effect of the $\Delta E_V$ at Si/HSL interface on the performance parameters - (a) $V_{oc}$, (b) $J_{sc}$, (c) $FF$, and (d) efficiency of CS Si solar cells for different $D_{it}$ ($cm^{-2}eV^{-1}$). Here both $V_{oc}$ and $J_{sc}$ is affected by the change in the passivation quality at the c-Si/HSL interface due to variation in $\Delta E_V$.

Finally, the solar cell efficiency (see Fig. 5d) also follows the trends of $V_{oc}$ and has its minimum at $\Delta E_c = 0eV$. The performance improves in both the directions, till over the barrier transport is affected and the efficiency becomes limited by $FF$ and $J_{sc}$. Accordingly, the best device performance is observed at $\Delta E_c \approx 0.4eV$.

*b) Effect of band discontinuity between HSL and c-Si:*
Figure 6 shows the effect of $\Delta E_V$ between c-Si and HSL in the presence of interface traps on solar cell performance parameters. As before, here we assume ideal conditions at ESL/c-Si interface (i.e., zero band offset and no traps). Fig. 6a and 6b shows the variation of $V_{oc}$ and $J_{sc}$, respectively, with $\Delta E_V$. While the change in $V_{oc}$ is very similar to that observed with ESL/c-Si band offset (see Section III(a)), surprisingly, there are significant differences in the $J_{sc}$ trends. We observe that the $J_{sc}$ varies with $\Delta E_V$ in contrast to the trends for ESL/Si interface (see Section IIIa). These puzzling trends are due to the distinct nature of ESL/c-Si and c-Si/HSL junctions. While the former is a PN junction, the latter is a PP$^+$ junction. Accordingly, the band bending in c-Si is more at the ESL/c-Si junction than the c-Si/HSL junction. This reduction in band bending increases the recombination loss at c-Si/HSL interface which reduces the collection efficiency of holes and hence the $J_{sc}$.

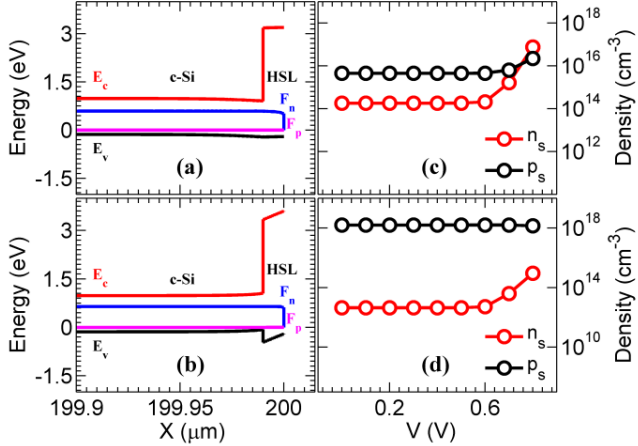

**Fig. 7**: The effect of $\Delta E_v$ on carrier densities near HSL. Parts (a, b) show the energy band diagram for $\Delta E_v = 0eV$ and $\Delta E_v = 0.4eV$ respectively at short circuit conditions. Parts (c, d) show the variation in interface carrier densities $n_s$ and $p_s$ with bias for $\Delta E_v = 0eV$ and $\Delta E_v = 0.4eV$ respectively.

Figure 7 shows the band diagram and carrier densities at Si/HSL junction for two different conditions. It is evident that there is negligible band bending in c-Si even for a large band offset of 0.4eV (compare Fig. 7a and 7b). Further, Fig. 7c shows that at short circuit conditions, the $n_s$ and $p_s$ differ by approximately 2 orders of magnitude for $\Delta E_v = 0eV$. However, for $\Delta E_v = 0.4eV$, $n_s$ and $p_s$ differ by 5 orders of magnitude at short circuit conditions (see part d). Thus the effect of interface recombination on $J_{sc}$ reduces with increase in $\Delta E_v$. Further the $J_{sc}$ decreases for $\Delta E_v > 0.5eV$ due to the reduction in efficiency of holes going over the barrier to the HSL. Fig. 6c and 6d shows the variation of $FF$ and efficiency with $\Delta E_v$. Due to the smaller band bending in c-Si near the HSL, the effect of interface traps is more at the c-Si/HSL interface compared to ESL/c-Si interface, as shown by the lower values of efficiency for the corresponding values of band discontinuity.

## IV. IMPLICATIONS

Our numerical simulations show that interface engineering is very crucial for Si based carrier selective solar cells. Indeed, for similar band offsets, the presence of traps is more detrimental at the selective layer that extracts the majority carrier. For example, Figs. 5 and 6 indicate that traps at HSL/c-Si interface (where holes, the majority carriers for p-substrate, are collected) reduce the efficiency quite significantly as compared to the same trap density at ESL/c-Si interface (where electrons, the minority carriers, are collected). Note that similar conclusions will hold good for n-type substrates as well, as the above effect is only influenced by the amount of band bending in c-Si. This has interesting implications on the choice of materials for Si based carrier selective solar cells. While the presence of a junction and hence the associated band bending allows considerable freedom in the choice of selective layer to extract the minority carriers, the selective layer for extraction of majority carriers should be chosen carefully with near perfect interface passivation properties. Accordingly, among the various choices, we speculate that a-Si based selective layers might be the most promising to extract the majority carrier (as a-Si could provide excellent interface passivation), while many other materials could be successful to extract the minority carriers.

Finally, our results indicate that the eventual performance of CS Si solar cells is dictated by the extent of band bending in c-Si. This effect is very similar to the field effect passivation of interface traps. We find that band offsets between CS material and c-Si play a significant role in reducing the interface recombination. As such many other parameters could also influence the band bending in Si. For example, the doping of CS layers can significantly affect the band bending in Si. Similarly, the dielectric constant of CS layers also has a non-intuitive effect on the band bending. Specifically, large doping or dielectric constant of CS results in an increased band bending in c-Si and hence are helpful to reduce interface recombination (also, as predicted by our analytical model, see Section II). These insights are also supported by detailed numerical simulations provided in appendix E (see Fig. 11). As such, band discontinuity, doping and dielectric constant of CS materials are the critical parameters that could affect the interface recombination and hence the efficiency of Si based carrier selective solar cells. Further, this information allows us to compare the promises of various CS materials like a:Si ($\Delta E_c \sim 0.3eV, \varepsilon = 11.9$), TiO$_2$ ($\Delta E_c \sim -0.05eV, \varepsilon \sim 85$), and ZnO ($\Delta E_c \sim -0.6eV, \varepsilon \sim 9$)[11], [16], [23] as ESL. For similar doping and $D_{it}$, our results indicate that a:Si might be the optimal choice and followed by ZnO. TiO$_2$ has the drawback of almost perfect band alignment; however has the advantage of large dielectric constant. These trends indicate that a-Si based carrier selective contacts could continue to yield the best performance as both the ESL and HSL (see section IIIb also) – a conclusion also partially supported by the excellent efficiencies achieved by HIT solar cells[24]. Future exploration of new CS materials can be immensely benefitted through a quantitative evaluation of material parameters (band offset, doping density, dielectric constant, and interface trap density) as detailed in this manuscript.

## V. CONCLUSIONS

To summarize, here we addressed the effect of CS material properties on the solar cell performance. We developed an analytical model to evaluate the effect of CS material parameters on the eventual efficiency and the same was validated using detailed numerical simulations. Curiously, we found that the optimal band alignment depends on the interface quality. If the interface quality is very good, then the efficiency is limited by over the barrier transport of carriers and hence small band offsets do not affect the performance. Otherwise, for not so ideal conditions at the interface, a band offset of around 0.2eV-0.4eV provides sufficient band bending to reduce the effect of interface recombination, thus improving the performance. In addition, our results show that the need for excellent interface passivation is more at the majority carrier extraction layer than at the minority carrier extraction layer. Further, we find that both the doping and the dielectric constant of the CS material have a similar effect on the performance. These interesting insights could be of broad interest to the community towards the selection of appropriate materials as carrier selective layers.

## VI. APPENDIX

**A. Parameters used in simulations:** To study the band offset effects, $\Delta E_c$ at the ESL/c-Si interface is varied from $-0.6eV$ to $+0.6eV$, while the barrier for holes is kept fixed (2.28eV). At the c-Si/HSL interface, $\Delta E_v$ is varied from $-0.2eV$ to $+0.5eV$, while the barrier for electrons is kept fixed (2.28eV). As a result, the band gap (and also the electron affinity) of ESL varies with the corresponding band offset used in each simulation. Accordingly, in our simulations the ESL band gap varies from 2.8eV to 4eV, which is comparable to the band gap of TiO$_2$ (~3.4eV). Similar arguments hold good for HSL as well. For ease of analysis, we have used same dielectric constant (6.215) for both ESL and HSL. However, many materials could have large dielectric constants (like TiO$_2$) and the effect of large dielectric constant is explored in appendix E (also mentioned in Section IV of main text). We consider uniform distribution of traps as the interface of CS layer and Si. The capture cross section of these traps was assumed as $10^{-16}$cm$^{-2}$. The rest of the parameters are provided in the table below.

| Parameter | c-Si | ESL | HSL |
|---|---|---|---|
| $N_c$(cm$^{-3}$) | $3.23 \times 10^{19}$ | $2.5 \times 10^{20}$ | $2.5 \times 10^{20}$ |
| $N_v$(cm$^{-3}$) | $1.83 \times 10^{19}$ | $2.5 \times 10^{20}$ | $2.5 \times 10^{20}$ |
| Mobility($cm^2V^{-1}s^{-1}$) (n, p)) | 1417, 470.5 | 20, 2 | 20, 2 |
| $\tau$ SRH (s) | $10^{-3}$ | $10^{-6}$ | $10^{-6}$ |
| Radiative Recombination coefficient ($cm^3s^{-1}$) | $1.1 \times 10^{-14}$[25] | | |
| Auger Coefficients ($cm^6s^{-1}$) (n,p) | $1 \times 10^{-31}$, $0.79 \times 10^{-31}$[26] | | |
| Doping(cm$^{-3}$)n/p | p - $10^{17}$ | n - $10^{17}$ | p - $10^{17}$ |

**Table. 1**: Parameters used in numerical simulations.

**B. Estimation of n$_s$ :** The interface carrier concentrations at the ESL/c-Si interface are estimated as follows. The electron density in c-Si at the ESL interface, n$_s$[27], is estimated using the formula,

$$n_s = \frac{n_i^2 e^{\frac{qV_{si}}{kT}}}{p_s}. \quad (11)$$

Here p$_s$, the hole density in c-Si at the ESL interface is estimated after finding the band bending in c-Si as given below using Poisson's equation. Fig. 8 shows the notations used for the depletion edges at ESL/c-Si interface.

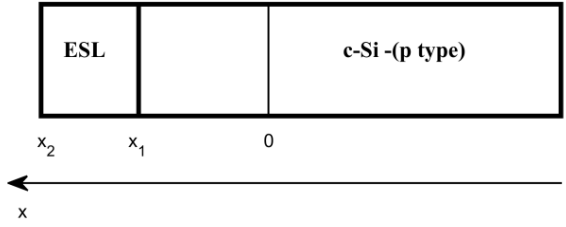

**Fig. 8:** Depletion region in CS Si solar cell at the ESL/c-Si interface. The edge of the depletion region in c-Si is denoted by x=0 while $x_1$ is the ESL/c-Si interface and $x_2$ in the edge of ESL.

Poisson's equation[27] is given by,

$$\nabla \xi = \frac{\rho}{\varepsilon}, \quad (12)$$

where $\xi$ is the electric field, $\rho$ is the charge density, and $\varepsilon$ is the permittivity.

For the depletion region in c-Si, the electric field follows the relation,

$$\xi_{si}(x) = \frac{\rho_{si} x}{\varepsilon_{si}} + c_1, \quad (13)$$

where $\xi_{si}$, $\rho_{si}$ and $\varepsilon_{si}$ are the electric field, the charge density and the permittivity, respectively in c-Si.

Since the electric field vanishes at the depletion edge, x=0, $c_1$ reduces to zero. For the depletion region in ESL, the electric field is given by,

$$\xi_{ESL}(x) = \frac{\rho_{ESL} x}{\varepsilon_{ESL}} + c_2, \quad (14)$$

where $\xi_{ESL}$, $\rho_{ESL}$ and $\varepsilon_{ESL}$ are the electric field, the charge density and the permittivity, respectively in ESL.

At $x = x_1$, the ESL/c-Si interface, using eq. (13) the electric field is determined as $\xi_{si}(x_1) = \frac{\rho_{si} x_1}{\varepsilon_{si}}$. Equating this to eq. (14) at $x = x_1$, $c_2$ is estimated as $c_2 = (\frac{\rho_{si}}{\varepsilon_{si}} - \frac{\rho_{ESL}}{\varepsilon_{ESL}})x_1$.

For depletion region in c-Si, the potential follows the relation

$$V_{si}(x) = \frac{-\rho_{si} x^2}{2\varepsilon_{si}}. \quad (15)$$

At $x = x_1$, the potential is given by,

$$V_{si}(x_1) = \frac{-\rho_{si} x_1^2}{2\varepsilon_{si}}. \quad (16)$$

For depletion region in ESL, the potential is given by,

$$V_{ESL}(x) = \frac{-\rho_{ESL} x^2}{2\varepsilon_{ESL}} - c_2 x. \quad (17)$$

At $x = x_2$, the potential is given by,

$$V_{ESL}(x_2) = \frac{-\rho_{ESL}(x_2^2 - x_1^2)}{2\varepsilon_{ESL}} - c_2(x_2 - x_1), \quad (18)$$

where, $x_2 = x_1 + w_{ESL}$.

Finally, the total potential in ESL and c-Si is related to applied bias by the relation,

$$V_{bi} - V_{app} = V_{eESL}(x_2) + V_{si}(x_1). \quad (19)$$

Using equations (16)-(19), the value of $x_1$ can be estimated and then using eq. (16) the voltage drop in c-Si is obtained. Using the voltage drop in c-Si $p_s$ is estimated as given below,

$$p_s = p_{bulk} e^{\frac{-qV_{si}}{kT}}, \quad (20)$$

where $p_{bulk}$ is the bulk majority carrier concentration in c-Si.

Finally eq. (11) is used to estimate the value of electron density at the interface.

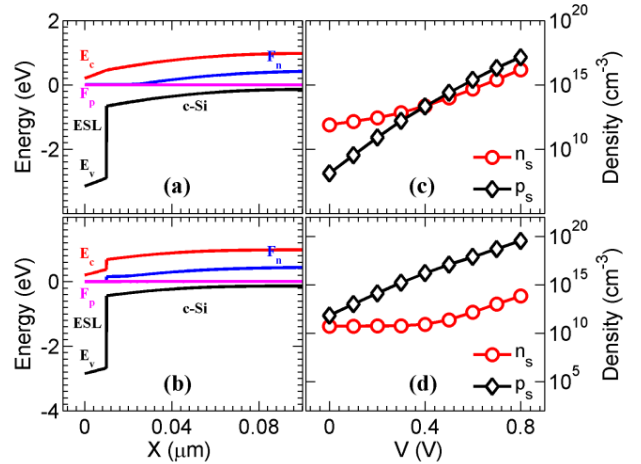

**Fig. 9:** The effect of negative values of $\Delta E_c$ on band bending and carrier densities at c-Si/ESL interface. Parts (a) and (b) show the energy band diagram for $\Delta E_c = 0 eV$ and $\Delta E_c = -0.3 eV$, respectively, at short circuit conditions. Parts (c, d) show the variation in interface carrier densities $n_s$ and $p_s$ with bias for $\Delta E_c = 0 eV$ and $\Delta E_c = -0.3 eV$, respectively.

**C. The effect of negative $\Delta E_c$ on device electrostatics:** Fig. 9a and 9b shows the energy band diagram near the ESL at short circuit conditions for $\Delta E_c = 0eV$ and $\Delta E_c = -0.3eV$, respectively. It is evident that the band bending in c-Si is more in the case for $\Delta E_c = 0eV$. Fig. 9c and 9d show the variation in the carrier densities with bias at the ESL/c-Si interface in the c-Si edge for $\Delta E_c = 0eV$ and $\Delta E_c = -0.3eV$, respectively. Due to the decrease in band bending for $\Delta E_c = -0.3eV$, the value of $n_s$ is less and the value of $p_s$ is more compared to the corresponding values for $\Delta E_c = 0eV$. As explained in section III(a) of the paper, this acts as field effect passivation and decreases the rate of trap assisted recombination at the interface, for large negative value of $\Delta E_c$. As a result, the $V_{oc}$ is expected to increase with negative value of $\Delta E_c$.

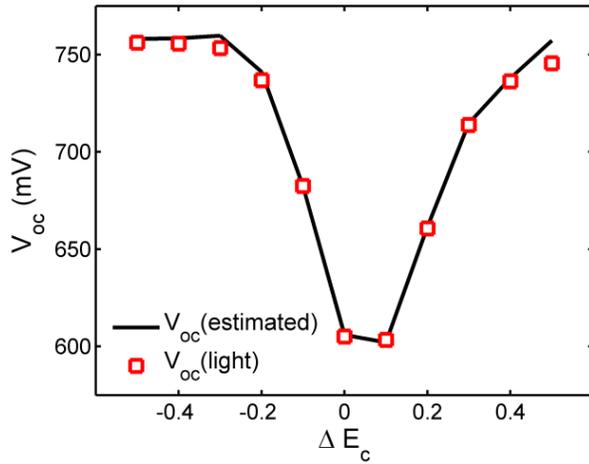

**Fig. 10:** Comparison of $V_{oc}$ obtained from simulated light IV characteristics (symbols) and estimated from dark IV using eq. 21 (solid line).

**D. The estimation of $V_{oc}$ from the dark characteristics:** The dark characteristics show that the ideality factor and $J_0$ are maximum at $\Delta E_c = 0eV$ and that both of them decrease in either direction from $\Delta E_c = 0eV$. Fig. 10 shows that a very good estimate of $V_{oc}$ is obtained using $J_0$ and ideality factor from dark characteristics. Here $V_{oc}$ was estimated using,

$$V_{oc} = \eta \frac{kT}{q} \ln \frac{J_{sc}}{J_0}, \quad (21)$$

where $\eta$ is the ideality factor.

**E. Effect of doping and dielectric constant in the SL:** Fig. 11 shows the effect of doping and dielectric in ESL with $\Delta E_c$ on the solar cell performance parameters. The effect of $\Delta E_c$ is explained in detail in Section III(a) of the paper. Fig. 11a shows the variation of $V_{oc}$ with $\Delta E_c$. It shows that $V_{oc}$ improves with the increase in doping and dielectric constant in the ESL for the corresponding positive values of $\Delta E_c$. The effect of increasing both the doping and the dielectric constant in the ESL is to increase the band bending in the c-Si region near the ESL. The increase in band bending improves the field effect passivation at the c-Si/ESL interface and hence improves the overall performance of the CS Si solar cell. Similar results are observed for HSL too.

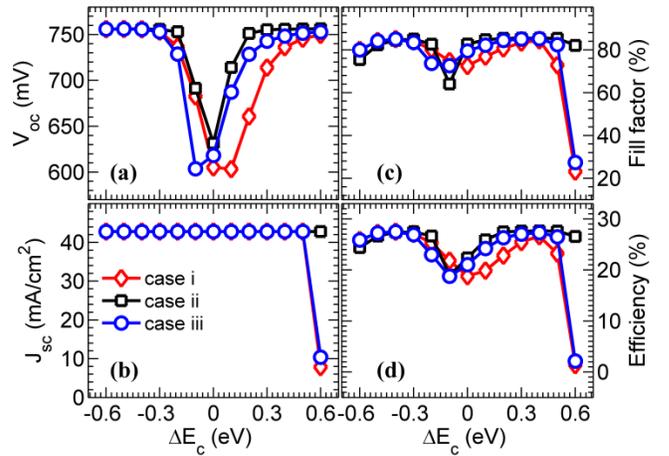

**Fig. 11:** The effect of the doping and dielectric constant in ESL on solar cell performance parameters for $D_{it} = 10^{12} cm^{-2} eV^{-1}$. Case (i) has a ESL doping of $10^{17} cm^{-3}$ and dielectric constant of 6.215, case (ii) has a ESL doping of $10^{17} cm^{-3}$ and dielectric constant of 85, and case (iii) has a ESL doping of $10^{18} cm^{-3}$ and dielectric constant of 6.215. The results indicate that the passivation quality increases with doping and dielectric constant for the same positive value of $\Delta E_c$.

**Acknowledgements:** This paper is based upon work supported in part by the Solar Energy Research Institute for India and the United States (SERIIUS), funded jointly by the U.S. Department of Energy (under Subcontract DE-AC36-08GO28308) and the Govt. of India's Department of Science and Technology (under Subcontract IUSSTF/JCERDC-SERIIUS/2012). The authors also acknowledge Center of Excellence in Nanoelectronics (CEN) and National Center for Photovoltaic Research and Education (NCPRE), IIT Bombay for computational facilities.